\documentclass[12pt]{article}
\usepackage{fullpage}

\title{Multi-Period Liability Clearing\\ via Convex Optimal Control}
\author{Shane Barratt \and Stephen Boyd}

\usepackage[colorlinks,allcolors=blue,bookmarks=false,hypertexnames=true]{hyperref}
\usepackage{mathtools,graphicx,graphics,psfrag,amsmath,amsfonts,verbatim,xcolor,color}
\usepackage{subcaption}

\newcommand{\BEAS}{\begin{eqnarray}}
\newcommand{\EEAS}{\end{eqnarray}}
\newcommand{\BEA}{\begin{eqnarray}}
\newcommand{\EEA}{\end{eqnarray}}
\newcommand{\BEQ}{\begin{equation}}
\newcommand{\EEQ}{\end{equation}}
\newcommand{\BIT}{\begin{itemize}}
\newcommand{\EIT}{\end{itemize}}
\newcommand{\BNUM}{\begin{enumerate}}
\newcommand{\ENUM}{\end{enumerate}}

\newcommand{\BA}{\begin{array}}
\newcommand{\EA}{\end{array}}

\newcommand{\eg}{{\it e.g.}}
\newcommand{\ie}{{\it i.e.}}

\newcommand{\ones}{\mathbf 1}

\newcommand{\reals}{{\mbox{\bf R}}}

\newcommand{\Tr}{\mathop{\bf Tr}}
\newcommand{\diag}{\mathop{\bf diag}}

\newcounter{algorithmctr}[section]
\renewcommand{\thealgorithmctr}{\thesection.\arabic{algorithmctr}}
   {\refstepcounter{algorithmctr}\begin{list}{}{%
       \setlength{\rightmargin}{0.03\linewidth}%
       \setlength{\leftmargin}{0.03\linewidth}}%
       \rmfamily\small
       \item[]{\setlength{\parskip}{0ex}\hrulefill\par%
        \nopagebreak{\bfseries\textsf{Algorithm \thealgorithmctr~}}}}%
   {{\setlength{\parskip}{-3ex}\nopagebreak\par\hrulefill} \end{list}}

\begin{document}
\maketitle

\begin{abstract}
We consider the problem of determining a sequence of payments among 
a set of entities that clear (if possible) the liabilities among them.
We formulate this as an optimal control problem, which is convex
when the objective function is, and therefore readily solved.
For this optimal control problem, we give a number of useful and interesting
convex costs and constraints that can be combined
in any way for different applications.
We describe a number of extensions, for example to handle
unknown changes in cash and liabilities,
to allow bailouts,
to find the minimum time to clear the liabilities,
or to minimize the number of non-cleared liabilities,
when fully clearing the liabilities is impossible.
\end{abstract}

\section{Introduction}

Large, complex networks of liabilities are
the foundation of modern financial systems.
According to the FDIC,
there were on the order of five thousand FDIC-insured banks
in the United States at the end of 2019 \cite{fdic}.
Each of these banks owe each other money
as a result of bank transfers, loans, and
securities issued.
Inter-bank settlement is handled today by simple
payment systems like Fedwire and CHIPS \cite[\S2.4]{bekaert2012international}.
Another example of complex liability networks
are derivatives exchanges and brokerages,
where there are liabilities between clients in the form of derivatives
contracts or borrowed shares.
A goal shared by all of the entities in these systems
is to clear or remove liabilities, which reduces risk and complexity.
Each system has its own goals and constraints
in its mission to clear liabilities, which must be accounted for.

We consider the general problem of liability clearing,
which is to determine a sequence of payments between a set of
financial entities to clear (if possible)
the liabilities among them over a finite time horizon.
We first observe that the dynamics in liability clearing are linear,
and describe methods that can be used to remove cycles of liability before
any payments are made.
We then formulate liability clearing as an optimal control problem
with convex objectives and constraints,
where the system's state is the cash held by each entity
and the liabilities
between each entity, and the input to the system is the payments
made by each entity to other entities.
This formulation has several benefits.
First, we can naturally incorporate
the goals and constraints in liability clearing in the stage cost function
of our optimal control problem.
Second, we can efficiently (globally) solve the problem
since it is convex.
Third, domain specific languages
for convex optimization make it easy to prototype
new liability clearing mechanisms.

We also extend our formulation to the case
where there are exogenous unknown inputs to the dynamics,
which represent uncertain future liabilities or cash flows.
We propose a solution method based on model predictive control,
or shrinking horizon control, which, at each time step, predicts the future
unknown inputs, plans a sequences of payments,
and then uses just the first of those payments
for the next time period.
We then illustrate our method on several simulated
numerical examples, comparing to a simple pro-rata payment baseline.
At the end of the paper, we discuss extensions and variations of our problem,
\eg, allowing bailouts, finding the minimum time
to clear a set of liabilities, non-time-separable costs,
infinite time liability control,
and how to minimize the number of non-cleared liabilities.

\paragraph{Outline.}
In \S\ref{s-related}
we discuss related work as well as its relation
to our paper.
In \S\ref{s-setup} we set out our notation,
and describe the dynamics equations and constraints.
In \S\ref{s-control} we formulate liability clearing
as a convex optimal control problem, and describe
a number of useful and interesting convex costs and constraints.
In \S\ref{s-mpc} we extend the optimal control formulation
to the case where the dynamics are subject to additional
uncontrollable exogenous terms, and propose a standard
method called model predictive control for this problem.
In \S\ref{s-examples} we illustrate the methods
described in this paper by applying them to
several numerical examples.
In \S\ref{s-extensions} we conclude
with a number of extensions and variations on our formulation.

\section{Related work}\label{s-related}
The liability clearing problem was originally proposed by
Eisenberg and Noe in 2001 \cite{eisenberg2001systemic}.
Their formulation involves determining a single
set of payments to be made between the entities,
in contrast to ours, which assumes a sequence of payments are made.
Their formulation assumes that these payments can be financed immediately
by payments received, whereas we make the realistic financing constraint
that entities cannot pay other entities more than the cash they
have on hand.
(This means that it may take multiple steps to clear liabilities.)
In this way, our formulation can be viewed as a supply chain
with cash as the commodity \cite[\S9.5]{boyd2018introduction}, while their formulation can be viewed
as a network flow problem \cite{harris1955fundamentals}, where cash can travel multiple steps
through the network at once.

They also make the assumption that all liabilities have equal priority
\cite[\S2.2]{eisenberg2001systemic},
\ie, each entity makes payments proportional to its liabilities
to the other entities (we call this a pro-rata constraint; see \S\ref{s-constraints}).
This means that instead of choosing a matrix of payments
between all the entities,
they only need to choose a vector of payments made by the entities;
the payments are then distributed according to the proportion of liability
(they call this vector the clearing vector).
Our formulation can include the constraint that payments are
made in proportion to liability, but we observe that
enforcing this constraint at each step is not
the most efficient strategy for clearing liabilities (see \S\ref{s-baseline}).
Because we can incorporate arbitrary (convex) costs and constraints, our formulation
is more flexible and realistic.

Eisenberg and Noe's original liability clearing formulation
has been extended in multiple ways
to include, \eg, default costs and rescue \cite{rogers2013failure},
cross-holdings \cite[\S2]{elsinger2009financial},
claims of different seniority \cite[\S6]{elsinger2009financial},
fire sales \cite{cifuentes2005liquidity},
multiple assets \cite{feinstein2019obligations},
and has been used to answer fundamental questions about contagion in
and shocks to large financial networks \cite{glasserman2015likely,feinstein2017measures}.
The extension of our methods to a couple of these cases
is described in \S\ref{s-extensions},
The sensitivity of clearing vectors to
liabilities has also been analyzed, using implicit differentiation
\cite{feinstein2018sensitivity} and the Farkas lemma \cite{khabazian2019vulnerability}.
Using the techniques of differentiable convex optimization
solution maps \cite{agrawal2019differentiable},
we can perform similar sensitivity analysis of our liability control problems.
Another tangential but related problem is
modeling of liquidity risk and funding runs;
of particular note here are the the Diamond and Dybvig model
of bank runs \cite{diamond1983bank} and the Allen and Gale
model of interbank lending \cite{allen2000financial}
(see, \eg, \cite[\S4]{glasserman2015likely} for a survey).

Eisenberg and Noe's formulation has also been extended
to multiple periods.
Capponi and Chen proposed a multi-period clearing framework
with a lender of last resort (\ie, bailouts, see \S\ref{s-bailouts})
and exogenous liabilities and cash flows (see \S\ref{sec:mpc}),
and proposed a number of heuristic policies
for controlling risk \cite{capponi2015systemic}.
However, their formulation does not include the financing constraint,
meaning liabilities can be (if possible) cleared in one step;
their focus is more on cash injection and defaults.
Other related works include
an extension to continuous time \cite{banerjee2018dynamic},
incorporation of multiple maturities and insolvency law \cite{kusnetsov2019interbank},
incorporation of contingent payments \cite{banerjee2019impact},
and an infinite-time treatment \cite{bardoscia2019full}.

\section{Notation and dynamics}\label{s-setup}
In this section we set out our notation,
and describe the dynamics equations and constraints.

\paragraph{Entities and cash held.}
We consider a financial system with
$n$ financial entities or agents, such as banks,
which we refer to as entities $1, \ldots, n$. 
These entities make payments to each other
over discrete time periods $t=1,\ldots,T$, where
$T$ is the time horizon.
The time periods could be any length of time,
\eg, each time period could represent a business day.
We let $c_t\in\reals_+^n$ denote
the cash held by each of the entities,
with $(c_t)_i$ being the amount held by entity $i$
in dollars at time period $t$.
If the entities are banks, then the cash held
is the bank's reserves, \ie, physical cash and
deposits at the central bank.
If the entities are individuals or corporations,
then the cash held is the amount of deposits at
their bank.

\paragraph{Liability matrix.}
Each entity has liabilities or obligations to the other entities,
which represent promised future payments.
We represent these liabilities
by the \emph{liability matrix} $L_t\in\reals_+^{n \times n}$,
where, at time period $t$,
$(L_t)_{ij}$ is the amount in dollars
that entity $i$ owes entity $j$ \cite[\S2.2]{eisenberg2001systemic}.
We will assume that $(L_t)_{ii}=0$, \ie, the entities do not owe anything to
themselves.
Note that $L_t \ones \in \reals_+^n$ is the vector of total liabilities of the entities, \ie,
$(L_t \ones)_i$ is the total amount that entity $i$ owes the other entities, in 
time period $t$,
where $\ones$ is the vector with all entries one.
Similarly, 
$L_t^T \ones \in \reals^n_+$ is the vector of total amounts owed to the entities by others,
\ie, $(L_t^T \ones)_i$ is the total amount owed to entity $i$ by the others.
The \emph{net liability} of the entities at time period $t$ is $L_t\ones - L_t^T\ones$,
\ie, $(L_t\ones-L_t^T\ones)_i$ is the net liability of entity $i$.
When $(L_t)_{ij}=0$, we say that the liability between entity
$i$ and $j$ is \emph{cleared} (in time period $t$).
The scalar quantity $\ones^T L_t \ones$ is the \emph{total gross liability}
between all the entities.  When it is zero, which occurs only when $L_t=0$, all
liabilities between the entities have been cleared.

\paragraph{Payment matrix.}
At each time step, each entity makes cash payments
to other entities.
We represent these payments by the \emph{payment matrix}
$P_t\in\reals_+^{n \times n}$,
$t=1,\ldots,T-1$,
where $(P_t)_{ij}$ is the amount in dollars
that entity $i$ pays entity $j$ in time period $t$.
We assume that $(P_t)_{ii}=0$, \ie, entities do not pay themselves.
Thus $P_t \ones \in \reals_+^n$ is the vector of total payments made by the entities to others
in time period $t$, \ie, $(P_t \ones)_i$ is the total cash paid by entity $i$ to the others.
The vector $P_t^T\ones \in \reals_+^n$ is the vector of total payments received by the entities 
from others in time period $t$, \ie, 
$(P_t^T\ones)_i$ is the total payment received by entity $i$ from the others.
Each entity can pay others no more than
the cash that it has on hand, so we have the constraint
\BEQ\label{e-cash-on-hand}
P_t\ones \leq c_t, \quad t=1,\ldots,T-1,
\EEQ
where the inequality is meant elementwise.

\paragraph{Dynamics.}
The liability and cash follow the linear dynamics
\begin{eqnarray}
L_{t+1} &=& L_t - P_t, \quad t=1,\ldots,T-1, \label{e-Lupdate}\\
c_{t+1} &=& c_t - P_t\ones + P_t^T\ones, \quad t=1,\ldots,T-1. \label{e-cupdate}
\end{eqnarray}
The first equation says that the liability
is reduced by the payments made,
and the second says that the cash is reduced by the total payments
and increased by the total payments received.

These dynamics can be extended to include an interest rate for cash, as well as 
additional cash flows into and out of the entities, and additional liabilities
among the entities.  For simplicity, we continue with the simple 
dynamics~(\ref{e-Lupdate}) and~(\ref{e-cupdate}) above, and describe some of these 
extensions in \S\ref{s-extensions}.

\paragraph{Monotonicity of liabilities.}
Since $L_t \geq 0$, these dynamics imply that
\BEQ\label{e-PtLt}
P_t \leq L_t, \quad t=1,\ldots,T-1,
\EEQ
\ie, each entity cannot pay another entity more than its liability.
We also observe that 
\BEQ\label{e-Lmonotone}
L_{t+1} \leq L_t, \quad t=1, \ldots, T-1,
\EEQ
where the inequality is elementwise, which means 
that each liability is non-increasing in time.
We conclude that if the liability of entity $i$ to entity $j$ is cleared in
time period $t$, it will remain cleared for all future time periods.
In other words, the sparsity pattern (\ie, which entries are nonzero)
of $L_t$ can only not increase over time.
The inequality \eqref{e-PtLt} implies that once a liability between entries
has cleared, no further payments will be made.
This tells us that the sparsity patterns of $P_t$ and $L_t$ are no larger than the 
sparsity pattern of $L_1$.

\paragraph{Net worth.}
The \emph{net worth} of each entity at the beginning of time period
$t$ is the cash it holds minus the total amount it owes others,
plus the total amount owed to it by others, or
\[
w_t = c_t - L_t \ones + L_t^T \ones,
\]
where $w_t \in \reals_+^n$ is the vector of net worth of the entities.
(The second and third terms are the negative net liability.)
The net worth is an invariant under the dynamics, since
\begin{eqnarray*}
w_{t+1} &=& c_{t+1} - L_{t+1} \ones + L_{t+1}^T \ones, \\
&=& c_t - P_t\ones + P_t^T\ones - (L_t - P_t) \ones + (L_t - P_t)^T \ones, \\
&=& c_t - L_t\ones + L_t^T\ones, \\
&=& w_t.
\end{eqnarray*}

\paragraph{Default.}
If $(w_1)_i < 0$, \ie, the initial net worth
of entity $i$ is negative,
then it will have to default;
it cannot reduce its net liability to zero.
If an entity defaults, then it
will find itself unable to fully pay the entities it owes money
to, which might cause those entities to default as well.
Such a situation is called a \emph{default cascade} \cite[\S2.4]{eisenberg2001systemic}.

\subsection{Liability cycle removal}

\paragraph{Graph interpretation.}
The liabilities between the entities can be interpreted
as a weighted directed graph, where the nodes
represent the entities, and the directed edges represent
liabilities between entities, with weights given by the liabilities.
In this interpretation, the liability matrix is simply the weighted adjacency matrix.

\paragraph{Liability cycle removal.}
Some of the liabilities between entities can be reduced or
removed without the need to make payments between them.
This happens when there are one or more \emph{liability cycles}.
A liability cycle is a cycle
in the graph described above,
or a sequence of positive liabilities that
starts and ends at the same entity and does not visit
an entity more than once.
If there is a liability cycle,
then each liability in the cycle can be
reduced by the smallest liability present in the cycle,
which reduces at least one of the liabilities in the cycle to zero (which therefore breaks the
cycle).
Removing a liability cycle in this manner keeps the net liabilities of
each entity,
$L_t\ones-L_t^T\ones$, constant.
The simplest case occurs with a cycle of length two: 
If $(L_t)_{ij}$ and $(L_t)_{ji}$ are both positive,
\ie, entities $i$ and $j$ each owe the other some positive amount,
then we can replace these liabilities with
\[
(L_t)_{ij}-\min\{(L_t)_{ij},(L_t)_{ji}\},
\qquad (L_t)_{ji}-\min\{(L_t)_{ij},(L_t)_{ji}\},
\]
which will reduce one of the two liabilities (the one that was originally smaller)
to zero.

Given a liability matrix $L$, we
give two ways to remove liability cycles,
a greedy algorithm and a formulation of
the problem as a linear program.
This problem is referred to in the literature
as portfolio compression \cite{marco2017compressing,schuldenzucker2019portfolio,veraart2019does}
and payment netting \cite{shapiro1978payments,o2014optimizing,o2017optimising}.

\paragraph{Greedy cycle clearing algorithm.}
The greedy cycle clearing algorithm begins by searching
for a liability cycle, which can be done using
a topological sort \cite{kahn1962topological}.
If there are no liability cycles, the algorithm terminates.
On the other hand, if there is a liability cycle, the
algorithm reduces each liability in the cycle by the smallest liability
present in the cycle, thus removing the cycle.
This process is repeated until there are no more liability cycles.
This algorithm was first proposed in 2009 in a patent filed by
TriOptima \cite{brouwer2009system}, a portfolio compression company
owned by the CME group that has reported clearing over 1000 trillion dollars of liabilities
through 2017.

\paragraph{Optimal cycle clearing via linear programming.}
The greedy algorithm described above can be improved upon
if our goal is not to just remove cycles,
but also to remove as much total gross liability as possible.
The problem is to find a new liability matrix $\tilde L\leq L$
with the smallest total gross liability,
subject to the constraint that the net liabilities remains the same.
This can be accomplished by solving the linear program
\BEQ
\begin{array}{ll}
\mbox{minimize} & \displaystyle \ones^T\tilde L \ones \\
\mbox{subject to} & L\ones - L^T\ones = \tilde L\ones - \tilde L^T \ones,\\
& 0 \leq \tilde L \leq L.
\end{array}
\EEQ
with variable $\tilde L$.

To the best knowledge of the authors, the linear programming formulation
of this problem was first proposed by Shapiro in 1978 \cite{shapiro1978payments}.
In his formulation, he incorporated transaction costs
by making the objective $\Tr(C^T(L-\tilde L))$ for a given transaction
cost matrix $C\in\reals_+^{n \times n}$.
Other objectives are possible, \eg, the sum of the squared liabilities \cite[\S3.3]{o2014optimizing}.

Liability cycle removal could be carried out before the payments
have begun.
From~(\ref{e-Lmonotone}), this implies that no cycles would appear; that is,
$L_t$ would contain no cycles for $t=1, \ldots, T$.
We note however that the methods described in this paper work
regardless of whether there are cycles, or whether liability cycle removal 
has been carried out; that is, liability cycle removal is optional for the methods
described in this paper.

\section{Liability control} \label{s-control}

\subsection{Optimal control formulation}
We now formulate the problem of finding a suitable sequence of payments
that clear (or at least reduce) the liabilities among the entities as a convex optimal
control problem.
Given an initial liability matrix $L^\mathrm{init}$
and cash $c^\mathrm{init}$,
the \emph{liability control problem} is
to choose a sequence of payments $P_1,\ldots,P_{T-1}$
so as to minimize a sum of stage costs,
\BEQ\label{e-cost}
\sum_{t=1}^{T-1} g_t(c_t,L_t,P_t) + g_T(c_T,L_T),
\EEQ
where the function
$g_t:\reals_+^n\times\reals_+^{n\times n}\times\reals_+^{n\times n}\to\reals\cup\{+\infty\}$
is the (possibly time-varying) stage cost,
and 
$g_T:\reals_+^n\times\reals_+^{n\times n}\to\reals\cup\{+\infty\}$
is the terminal stage cost.

Infinite values of the stage cost $g_t$ (or $g_T$) are used to express 
constraints on $c_t$, $L_t$, or $P_t$.
To impose the constraint $(c_t,L_t,P_t)\in\mathcal C_t$,
we define $g_t(c_t,L_t,P_t)=+\infty$ for $(c_t,L_t,P_t)\not\in\mathcal C_t$.
As a simple example, the final stage cost 
\[
g_T(c_T,L_T) = \left\{ \begin{array}{ll} 0 & L_T=0, \\ \infty & L_T \neq 0, \end{array}\right.
\]
imposes the constraint that the sequence of payments must result in all liabilities 
cleared at the end of the time horizon.
Here $g_T$ is the \emph{indicator function} of the constraint $L_T=0$.
(The indicator function of a constraint has the value $0$
when the constraint is satisfied, and $+\infty$ when it is violated.)

The liability control problem has the form
\BEQ
\begin{array}{ll}
\mbox{minimize} & \sum_{t=1}^{T-1} g_t(c_t,L_t,P_t)  + g_T(c_T,L_T)\\
\mbox{subject to} & L_{t+1} = L_t - P_t, \quad t=1,\ldots,T-1,\\
& c_{t+1} = c_t - P_t\ones + P_t^T\ones, \quad t=1,\ldots,T-1,\\
& P_t\ones \leq c_t, \quad t=1,\ldots,T-1,\\
& P_t \geq 0, \quad t=1,\ldots,T-1,\\
& L_t \geq 0, \quad t=1,\ldots,T,\\
& c_1 = c^\mathrm{init}, \quad L_1 = L^\mathrm{init}, \quad c_T \geq 0,
\end{array}
\label{eq:liability_control}
\EEQ
with variables $c_t$, $L_t$, $t=1, \ldots, T$, and $P_t$, $t=1,\ldots, T-1$.
(The constraint $c_t \geq 0$ is implied by $P_t\ones\leq c_t$.)
We refer to this as the \emph{liability clearing control problem}.
It is specified by the stage cost functions $g_1, \ldots, g_T$,
the initial liability matrix $L^\mathrm{init}$,
and the initial cash vector $c^\mathrm{init}$.
We observe that the last four sets of inequality constraints could be absorbed into
the stage cost functions $g_t$ and $g_T$;
for clarity we include them in \eqref{eq:liability_control} explicitly.

\paragraph{Convexity.}
We will make the assumption that the stage cost functions $g_t$ and $g_T$
are convex, which implies that the liability control
problem~(\ref{eq:liability_control}) is a convex optimization problem~\cite{boyd2004convex}.
This implies that it can be (globally) solved efficiently, even at large scale;
this is discussed further in~\S\ref{s-solving}.
Perhaps more important from a practical point of view is that it can be solved
with near total reliability, with no human intervention, and at high speed if 
needed.

We make the assumption not just because of the computational advantages that convexity
confers, but also because there are very reasonable choices of the cost 
functions that satisfy the convexity assumption.
It is also true that some reasonable cost functions are not convex;
we give an example in \S\ref{sec:num_entitites}.

\subsection{Constraints}
\label{s-constraints}
In this section we describe some examples of useful constraints,
which can be combined with each other or any of the cost functions
described below.  They are all convex.

\paragraph{Liability clearance.}
We can constrain the liabilities to be fully cleared
at time $T$ with the constraint 
\[
L_T = 0.
\]
If $w_1\not\geq 0$ or
the liabilities cannot be cleared in time,
the liability clearing problem \eqref{eq:liability_control}
with this constraint will be infeasible.

\paragraph{Pro-rata constraint.}
The proportional liability of each entity
is the proportion of its total liability
that it owes to the other entities, which
for entity $i$ is $(L_t)_i/(L_t\ones)_i$.
We can constrain
the final proportional liability of each entity
to be equal to the initial proportional liability
with the linear constraint
\[
\diag(L_1\ones) L_T = \diag(L_T\ones) L_1,
\]
where $\diag(x)$ is the diagonal matrix
with $x$ on its diagonal.
This constraint also holds if $L_T=0$, \ie, the sequence of payments
clears all liabilities.  

\paragraph{Cash minimums.}
Cash minimums, represented by the vector $c^\mathrm{min} \in \reals_+^n$,
where $(c^\mathrm{min})_i$ is the minimum cash that the entity $i$ is allowed to hold,
can be enforced with the constraint
\[
c_t \geq c^\mathrm{min}, \quad t=1,\ldots,T.
\]
Cash minimums can arise for a number of reasons,
one of them being reserve requirements for
banks \cite{FEDreserverequirements}.

\paragraph{Payment maximums.}
We can constrain the payment between entities
to be below some maximum payment $P^\mathrm{max}\in\reals_+^{n\times n}$,
where $(P^\mathrm{max})_{ij}$ is the maximum allowable payment from
entity $i$ to entity $j$, with the constraint
\[
P_t \leq P^\mathrm{max}, \quad t=1,\ldots,T-1.
\]
We can impose a limit on how much cash
each entity uses for payments with the constraint
\[P_t\ones \leq \beta c_t, \quad t=1,\ldots,T-1,\]
where $0<\beta \leq 1$ is the fraction of the entity's cash that can
be used to make payments in each time period.

\paragraph{Payment deadlines.}
Deadlines on payments
are represented by the set
\[
\Omega \subseteq \{1,\ldots,T\} \times \{1,\ldots,n\} \times \{1,\ldots,n\}.
\]
If $(t,i,j)\in\Omega$,
we require that the liability between entities $i$ and $j$
becomes zero at time $t$.
This results in the constraints
$(L_t)_{ij} = 0$ for all $(t,i,j) \in \Omega$.

\paragraph{Progress milestones.}
We can impose the constraint that the liabilities are reduced by
the fraction $\eta\in(0,1)$ in $\tau$ time 
periods, with $L_\tau \leq \eta L_1$.

\subsection{Costs}
In this section we list some interesting and useful convex stage costs.
We note that any combination of the constraints above can be included with
any combination of the costs listed below, by adding their indicator functions
to the cost. 

\paragraph{Weighted total gross liability.}
A simple and useful stage cost is
a weighted total gross liability,
\BEQ \label{e-total-liability-cost}
g_t(c_t,L_t,P_t) = \Tr(C^TL_t) = \sum_{i=1}^n \sum_{j=1}^n C_{ij}(L_t)_{ij},
\EEQ
where the matrix $C\in\reals_+^{n\times n}$ represents
the (marginal) cost of each liability.
When $C=\ones\ones^T$ (\ie, $C_{ij}=1$ for all $i$ and $j$),
this stage cost is simply the total
gross liability $\ones^T L_t \ones$ at time $t$.
When $C$ is not the all ones matrix, it encourages
reducing liabilities $L_{ij}$ with higher weights $C_{ij}$.

\paragraph{Total squared gross payment.}
Another simple and useful stage cost is the total squared gross
payment,
\[
g_t(c_t,L_t,P_t) = \Tr(D^TP_t^2) = \sum_{i=1}^n \sum_{j=1}^n D_{ij}(P_t)_{ij}^2,
\]
where $D\in\reals_+^{n \times n}$ represents
the cost of each squared payment,
and the square is taken elementwise.
This stage cost is meant to reduce
the size of payments made between entities.
As a result of the super-linearity of the square function,
it is more sensitive to large payments between
the entities than smaller ones.
In control terms, the sum of squared payments
is our \emph{control effort}, which we would like to be small.
It is a traditional term in optimal control.

\paragraph{Distance from cash to net worth.}
If the liability is cleared, \ie, $L_t=0$, then
the cash held by each entity will be equal to its net worth, or $c_t=w_t$.
We can penalize the distance from the cash held by each
entity to its net worth with, \eg, the cost function
\[
g_t(c_t,L_t,P_t) = \|c_t - w_1\|_2^2.
\]
If we want to make $c_t$ exactly equal to $w_1$ in as many
entries as possible as quickly as possible, we can replace the cost above
with the $\ell_1$ norm $\|c_t - w_1\|_1$.

\paragraph{Time-weighted stage cost.}
Any of these stage costs can be time-weighted.
That is, if the stage cost is time-invariant,
\ie, $g_t=g$ for some stage cost $g$,
the time-weighted stage cost is
\[
g_t(c_t,L_t,P_t) = \gamma^{t-1} g(c_t,L_t,P_t),
\]
where $\gamma >0$.  For $\gamma>1$, this
stage cost preferentially rewards the stage cost being
decreased later (\ie, for large $t$);
for $\gamma < 1$, it represents a traditional discount factor, which
preferentially rewards the stage cost being decreased earlier (\ie, for small $t$).
With $\gamma=1$, we treat stage costs at different time periods the same.

\subsection{Computational efficiency}
\label{s-solving}
Since problem \eqref{eq:liability_control} is a convex optimization
problem, it can be solved efficiently \cite{boyd2004convex},
even for very large problem sizes.
The number of variables and constraints
in the problem is on the order $Tn^2$.
However, this convex optimization problem is often very sparse.
The inequalities \eqref{e-PtLt} and \eqref{e-Lmonotone} imply that $L_t$ and $P_t$
can only have nonzero entries where $L^\mathrm{init}$ does.
This means that the number of variables can be reduced to
order $T\mathbf{nnz}(L^\mathrm{init})$ variables, where
$\mathbf{nnz}(L^\mathrm{init})$ is the number of nonzero entries in the initial
liability matrix.
(In appendix \ref{s-form2}, we give an alternative formulation of
the liability clearing control problem that exploits this sparsity preserving 
property.)
Due to the block-banded nature of the optimal control problem,
the computational complexity grows linearly in $T$; see, \eg,
\cite[\S A.3]{boyd2004convex}.

As a practical matter, we can easily solve the liability clearing 
problem with $n=1000$ entities,
$\mathbf{nnz}(L^\mathrm{init}) = 5000$, and $T=20$, using generic methods running 
on an Intel i7-8700K CPU, in under a minute.
Small problems, with say $n=10$ entities,
$\mathbf{nnz}(L^\mathrm{init}) = 30$, and $T=20$ can be solved 
in under a millisecond, using techniques of code generation such as CVXGEN \cite{mattingley2009automatic,mattingley2012cvxgen}.

It is very easy to express the liability clearing control problem using 
domain specific languages for convex optimization, such as CVX \cite{grant2008cvx,grant2014cvx},
YALMIP \cite{lofberg2004yalmip}, CVXPY \cite{diamond2016cvxpy,agrawal2018cvxpy}, Convex.jl
\cite{udell2014convexjl}, and CVXR \cite{fu2019cvxr}.  
These languages make it easy to 
rapidly prototype and experiment with different cost functions and constraints.
In each of these languages, the liability control
problem can be specified in just a few tens of lines
of very clear and transparent code.

\subsection{Pro-rata baseline method}
\label{s-baseline}
We describe here a simple and intuitive scheme for determining
cash payments $P_1, \ldots, P_{T-1}$.  We will use this as a baseline method
to compare against the optimal control method described above.

The payment $P_t$ is determined as follows.
At each time step, each
entity pays as much as possible
pro-rata, \ie, in proportion to how much it owes the other
entities, up to its liability.
Define the liability proportion matrix as
\[
\Pi = \diag(1/(L^\mathrm{init}\ones)) L^\mathrm{init},
\]
so $(\Pi)_{ij}$ is the fraction
of entity $i$'s total liability
that it owes to entity $j$.
The pro-rata baseline has the form
\BEQ\label{e-pro-rata}
P_t = \min(\diag(c_t)\Pi, L_t),
\EEQ
where $\min$ is taken elementwise.
We will see that the (seemingly sensible) pro-rata baseline
is not an efficient strategy for optimally clearing liabilities.

\section{Liability control with exogenous unknown inputs}\label{s-mpc}

In this section we extend the optimal control
formulation in \S\ref{s-control} to handle additional (exogenous) 
terms in the liability and cash dynamics, unrelated to
the clearing process and payments.
When these additional terms are known, we obtain a straightforward
generalization of the liability clearing control problem, with a few extra
terms in the dynamics equations.
For the case when they are not known ahead of time,
we propose a standard method called model predictive control (MPC), or 
shrinking horizon control \cite{bemporad2006model,rawlings2009model,mattingley2011receding}.
MPC has been used successfully in a wide variety of applications,
for example, in supply chain management \cite{cho2003supply},
finance \cite{boyd2017multi},
automatic control \cite{falcone2007predictive,blackmore2010minimum},
and energy management \cite{ma2011model,soltani2011load,moehle2019dynamic}.
It has been observed to work well even when the forecasts are not
particularly good \cite[\S4]{wang2009performance}.

\subsection{Optimal control with exogenous inputs}
We replace the dynamics equations \eqref{e-Lupdate} and \eqref{e-cupdate} with
\begin{eqnarray}
L_{t+1} &=& L_t - P_t + W_t, \quad t=1,\ldots,T-1, \label{e-Lupdate-random}\\
c_{t+1} &=& c_t - P_t\ones + P_t^T\ones + w_t, \quad t=1,\ldots,T-1, \label{e-cupdate-random}
\end{eqnarray}
where $W_t\in\reals^{n\times n}$ is the liability adjustment at time $t$,
and $w_t\in\reals^n$ is the exogenous cash flow at time $t$.
The liability adjustment $W_t$ can originate
from entities creating new liability agreements;
the cash flow $w_t$ can originate from payments received or made by an entity,
unrelated to clearing liabilities.
The terms $W_t$ and $w_t$ are exogenous inputs in our dynamics, \ie, additional
terms that affect the liabilities and cash, but are outside our control (at least, for
the problem of clearing liabilities).
The cash on hand constraint \eqref{e-cash-on-hand} is modified to be
\BEQ\label{e-cash-on-hand-exog}
P_t\ones \leq c_t + w_t, \quad t=1,\ldots,T-1,
\EEQ
where $c_t + w_t$ is the cash on hand after the exogenous cash flow.

When the exogenous inputs are known (which might occur, for example,
when all the exogenous cash flows and liability updates are planned or scheduled), 
we obtain a straightforward generalization of the liability clearing control problem,
\BEQ\label{e-exog-prob}
\begin{array}{ll}
\mbox{minimize} & \sum_{t=1}^{T-1} g_t(c_t,L_t,P_t)  + g_T(c_T,L_T)\\
\mbox{subject to} & L_{t+1} = L_t - P_t + W_t, \quad t=1,\ldots,T-1,\\
& c_{t+1} = c_t - P_t\ones + P_t^T\ones + w_t, \quad t=1,\ldots,T-1,\\
& P_t\ones \leq c_t + w_t, \quad t=1,\ldots,T-1,\\
& P_t \geq 0, \quad t=1, \ldots, T-1,\\
& c_t \geq 0, \quad L_t \geq 0, \quad t=1, \ldots, T,\\
& c_1 = c^\mathrm{init}, \quad L_1 = L^\mathrm{init},
\end{array}
\EEQ
with variables $c_t$, $L_t$, and $P_t$.

\subsection{Optimal control with unknown exogenous inputs}
\label{sec:mpc}

We now consider a more common case, where $w_t$ and $W_t$ are not known, or not fully
known, when the sequence of payments is chosen.
It would be impossible to choose the payment in time period $t$
without knowing $w_t$; otherwise we cannot be sure to satisfy \eqref{e-cash-on-hand-exog}.
For this reason we assume that $W_t$ and $w_t$ are known at time period $t$, 
and therefore can be used when we choose the payment $P_t$.
(An alternative interpretation is that the exogenous cash arrives before we
make payments in period $t$.)
Thus at time period $t$, when $P_t$ is chosen, we assume that $w_1, \ldots, w_t$
and $W_1, \ldots, W_t$ are all known.

\paragraph{Forecasts.} At time period $t$, we do not know $w_{t+1}, \ldots, w_{T-1}$ or
$W_{t+1}, \ldots, W_{T-1}$.
Instead we use \emph{forecasts} of these quantities, which we denote by
\[
\hat w_{\tau \mid t}, \quad
\hat W_{\tau\mid t}, \quad \tau=t+1,\ldots,T-1.
\]
We interpret the subscript $\tau|t$ as meaning our forecast of the quantity 
at time period $\tau$, made at time period $t$.
These forecasts can range from sophisticated ones based on machine learning to 
very simple ones, like 
$\hat w_{\tau \mid t} =0$, $\hat W_{\tau \mid t} =0$, 
\ie, we predict that there will be no future adjustments to the cash or liabilities.
We will take $\hat w_{\tau|t} = w_\tau$ and
$\hat W_{\tau|t} = W_\tau$ for $\tau \leq t$; that is, our `forecasts'
for the current and earlier times are simply the values that were observed.

\paragraph{Shrinking horizon policy.}
We now describe a common heuristic for choosing $P_t$ at time period $t$,
called MPC.
The idea is very simple: we solve the problem \eqref{e-exog-prob}, over the remaining
horizon from time periods $t$ to $T$, replacing the unknown quantities with
forecasts.
That is, we solve the problem
\BEQ\label{e-mpc}
\begin{array}{ll}
\mbox{minimize} & \sum_{\tau=t}^{T-1} g_\tau(c_\tau,L_\tau,P_\tau)  + g_T(c_T,L_T)\\
\mbox{subject to} & L_{\tau+1} = L_\tau - P_\tau + \hat W_{\tau|t}, \quad \tau=t,\ldots,T-1,\\
& c_{\tau+1} = c_\tau - P_\tau\ones + P_\tau^T\ones + \hat w_{\tau|t}, \quad \tau=t,\ldots,T-1,\\
& P_\tau\ones \leq c_\tau +w_\tau, \quad \tau=t,\ldots,T-1,\\
& P_\tau \geq 0, \quad \tau=t, \ldots, T-1,\\
& c_\tau \geq 0, \quad L_\tau \geq 0, \quad \tau=t, \ldots, T,
\end{array}
\EEQ
with variables $c_{t+1}, \ldots, c_T$, $L_{t+1}, \ldots , L_T$, and $P_t, \ldots, P_{T-1}$.
In \eqref{e-mpc}, $c_t$ and $L_t$ are known; they are not variables,
and we take $\hat W_{t|t} = W_t$ and $\hat w_{t|t} = w_t$, which are known.
We can interpret the solution of \eqref{e-mpc} as a \emph{plan of action}
from time period $t$ to $T$.

We choose $P_t$ as the value of $P_t$ that is a solution of \eqref{e-mpc}.
Thus, at time period $t$ we \emph{plan} a sequence of payments (by solving \eqref{e-mpc});
then we \emph{act} by actually making the payments in the first step of our plan.
MPC has been observed to perform well in many applications, even when
the forecasts are not particularly good, or simplistic (\eg, zero).

\paragraph{Pro-rata baseline policy.}
We observe that the pro-rata baseline payments \eqref{e-pro-rata}
are readily extended to the case when we have exogenous inputs, with $w_t$ and $W_t$
known at time period $t$.
First, we define the liability proportion matrix at time $t$ as
\[
\Pi_t = \diag(1/L^\mathrm{run}_t) L^\mathrm{run}_t,
\]
where $L^\mathrm{run}_t = L^\mathrm{init} + \sum_{\tau=1}^t W_\tau$
is the running sum of liabilities.
The pro-rata baseline policy then has the form
\BEQ\label{e-pro-rata-exog}
P_t = \min(\diag(c_t+w_t)\Pi_t, L_t+W_t).
\EEQ

\section{Examples}\label{s-examples}

The code for all of these examples
has been made available online at
\[
\verb|www.github.com/cvxgrp/multi_period_liability_clearing|.
\]
We use CVXPY \cite{diamond2016cvxpy,agrawal2018cvxpy} to formulate the problems
and solve them with MOSEK \cite{aps2020mosek}.

\paragraph{Initial liability matrix.}
We use the same initial liability matrix $L^\mathrm{init}$
for each example, with $n=200$ entities.
We choose the sparsity pattern of $L^\mathrm{init}$ as
$2000$ random off-diagonal entries (so on average, each entity 
has an initial liability to 10 others).
The nonzero entries of $L^\mathrm{init}$
are then sampled independently from a standard log-normal
distribution.
While we report results below for this one problem instance,
numerical experiments with a wide variety of other instances
show that the results are qualitatively similar.
We note that our example is purely illustrative,
and that further experimentation needs to be performed
on problem instances that bear more structural similarity to real
world financial networks \cite{boss2004network}.

\subsection{Liability clearing}
\label{sec:liability_clearing_example}
We consider the problem of clearing liabilities
over $T=10$ time steps,
\ie, we have the constraint that the final liabilities
are cleared, $L_T=0$.
We set the initial cash to the minimum nonnegative
cash required so each entity has nonnegative net worth,
or
\[
c^\mathrm{init} = \max(L^\mathrm{init}\ones - (L^\mathrm{init})^T\ones, 0),
\]
where $\max$ is meant elementwise.

\paragraph{Total gross liability.}
\begin{figure}
    \centering %
    \begin{subfigure}{0.45\textwidth}
    \centering
  \includegraphics[width=\linewidth]{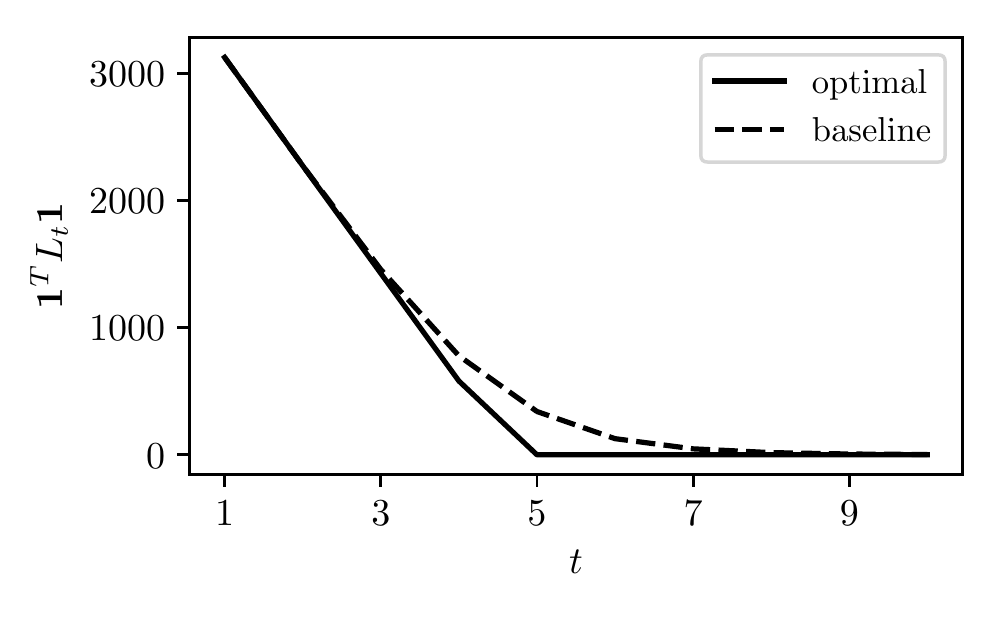}
  \caption{Total gross liability.}
\end{subfigure} %
\begin{subfigure}{0.45\textwidth}
\centering
  \includegraphics[width=\linewidth]{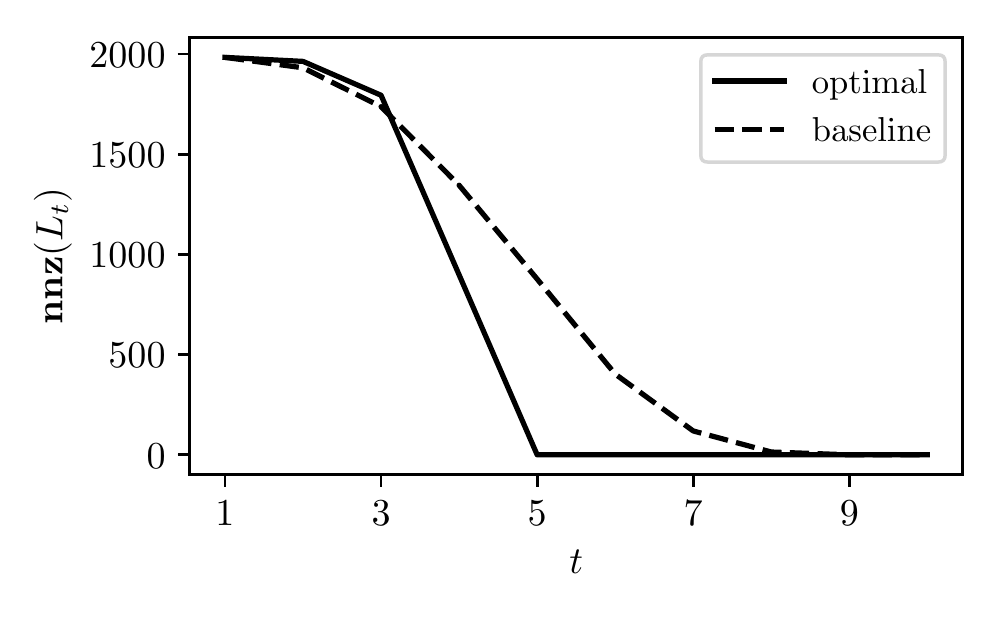}
  \caption{Number of non-cleared liabilities.}
\end{subfigure} %
\caption{Minimizing the sum of total gross liabilities.
The solid line is the optimal payment schedule. The dashed line
is the pro-rata baseline method.}
\label{fig:1}
\end{figure}
The first stage cost function we consider is
\[
g_t(c_t,L_t,P_t) = \ones^TL_t\ones,
\]
the total gross liability at each time $t$.
We compare the solution to the liability control
problem \eqref{eq:liability_control} using this stage cost function
with the pro-rata baseline method described in \S\ref{s-baseline}.
The total gross liability and the number of non-cleared liabilities
at each step of both sequences of payments are shown in figure \ref{fig:1}.
The optimal sequence of payments clears the liabilities by $t=5$, while
the baseline clears them by $t=8$.

\paragraph{Risk-weighted liability.}
\begin{figure}
    \centering %
    \begin{subfigure}{0.45\textwidth}
    \centering
  \includegraphics[width=\linewidth]{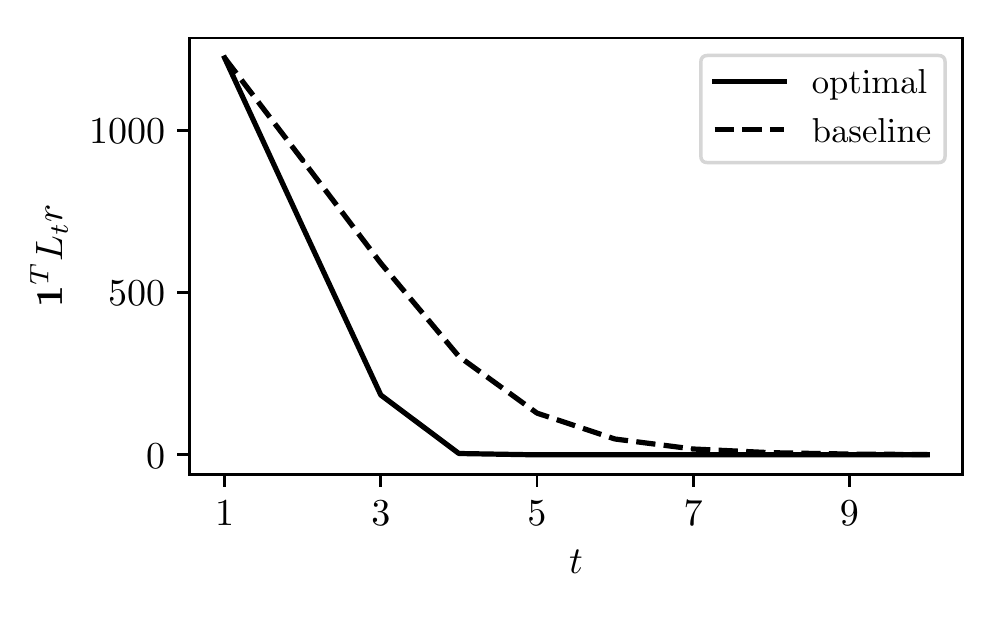}
  \caption{Risk-weighted liability.}
\end{subfigure} %
\begin{subfigure}{0.45\textwidth}
\centering
  \includegraphics[width=\linewidth]{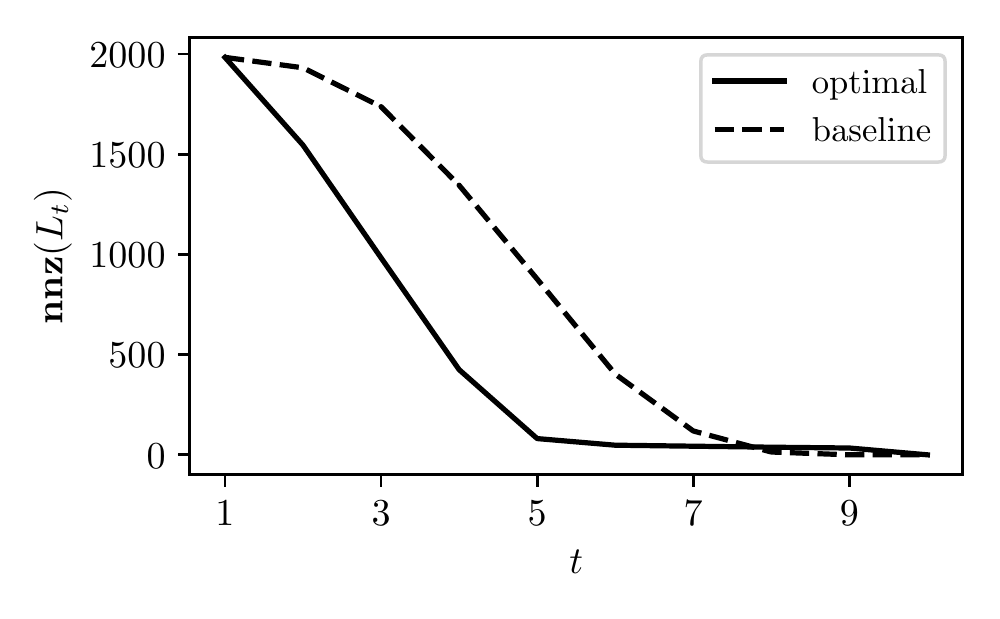}
  \caption{Number of non-cleared liabilities.}
\end{subfigure} %
\caption{Minimizing the sum of risk-weighted liabilities.
The solid line is the optimal payment schedule. The dashed line
is the pro-rata baseline method.}
\label{fig:2}
\end{figure}
Suppose we believe that the risk of each entity
is proportional to $r=\exp(-w_1)$, where $\exp$
is taken elementwise,
\ie, higher net worth implies lower risk.
A reasonable stage cost function is then
risk-weighted liability
\[
g_t(c_t,L_t,P_t)=\ones^T L_t r.
\]
This stage cost encourages clearing the liabilities
for high risk entities before low risk entities.
We compare the solution to the liability control
problem \eqref{eq:liability_control} using this stage cost function
with the pro-rata baseline method in \S\ref{s-baseline}.
The total gross liability and the number of non-cleared liabilities
at each step of both sequences of payments are shown in figure \ref{fig:2}.
We observe that the liabilities are still cleared by $t=5$,
but the liabilities are much sparser,
since the liabilities of high risk entities are cleared before
those of low risk entities.
We also note that the optimal payment sequence is much
faster at reducing risk than the baseline.

\paragraph{Total squared gross payment.}
\begin{figure}
    \centering %
    \begin{subfigure}{0.45\textwidth}
    \centering
  \includegraphics[width=\linewidth]{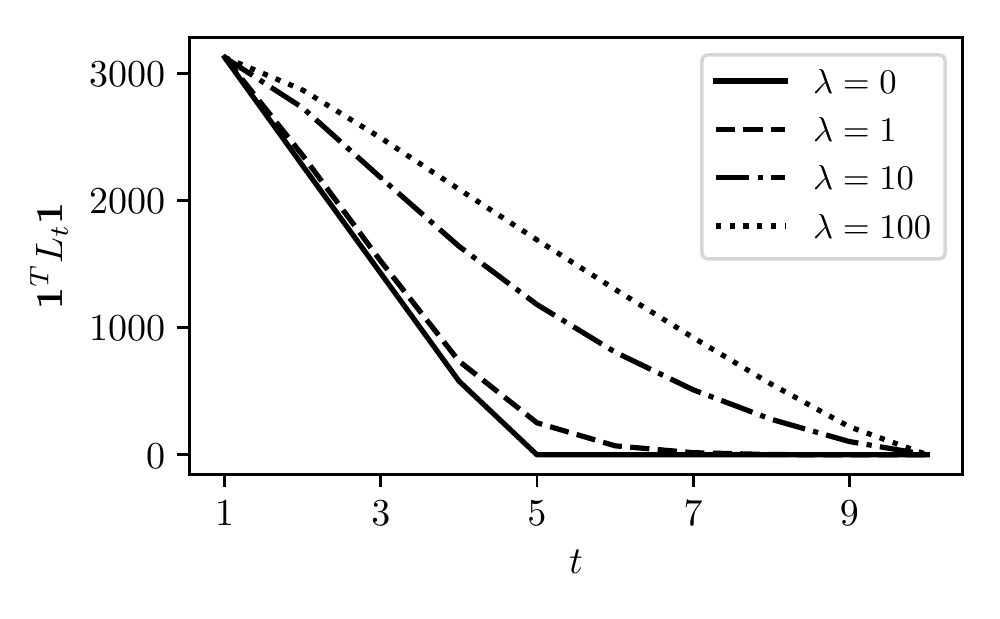}
  \caption{Total gross liability.}
\end{subfigure} %
\begin{subfigure}{0.45\textwidth}
\centering
  \includegraphics[width=\linewidth]{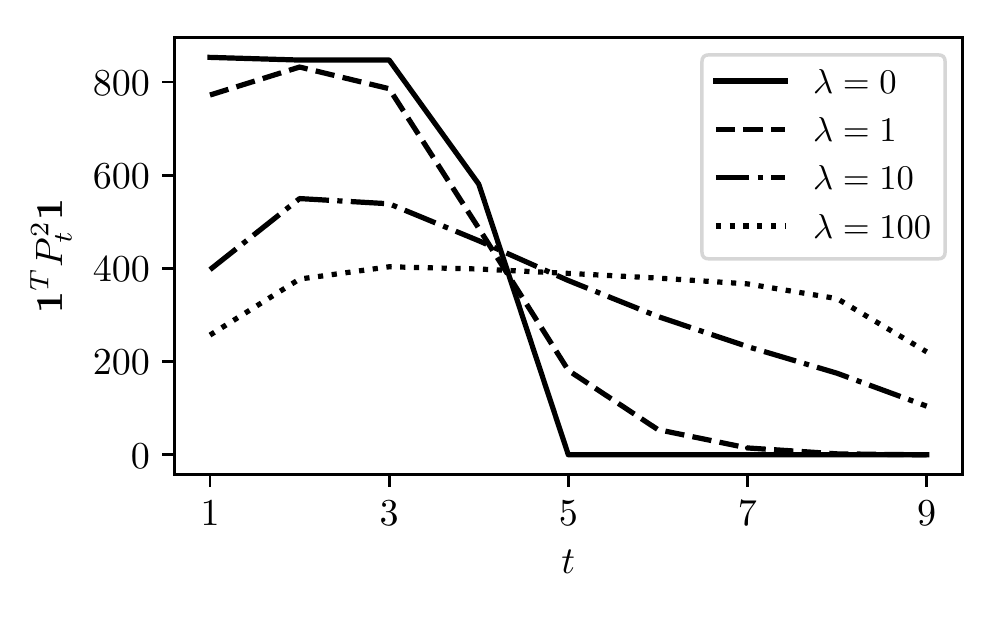}
  \caption{Total squared gross payment.}
\end{subfigure} %
\caption{Minimizing the sum of total gross liability plus total
squared gross payment for various values of $\lambda$.}
\label{fig:3}
\end{figure}
To the total gross liability stage cost above, we add the total squared payments,
resulting in the stage cost
\[
g_t(c_t,L_t,P_t)=\ones^T L_t \ones + \lambda \ones^T P_t^2 \ones,
\]
where $\lambda > 0$ is a parameter.
This choice of stage cost penalizes
large payments, and stretches the liability clearing
over a longer period of time. (We retain, however, the liability clearing
constraint $L_T=0$.)
We plot the optimal total gross liability and
the total squared gross payment for various values of $\lambda$
in figure \ref{fig:3}.
(We do not compare to the pro-rata baseline because it does not
seek to make payments small.)

\subsection{Liability reduction}
\label{sec:liability_reduction}

\begin{figure}
    \centering %
    \begin{subfigure}{0.45\textwidth}
    \centering
  \includegraphics[width=\linewidth]{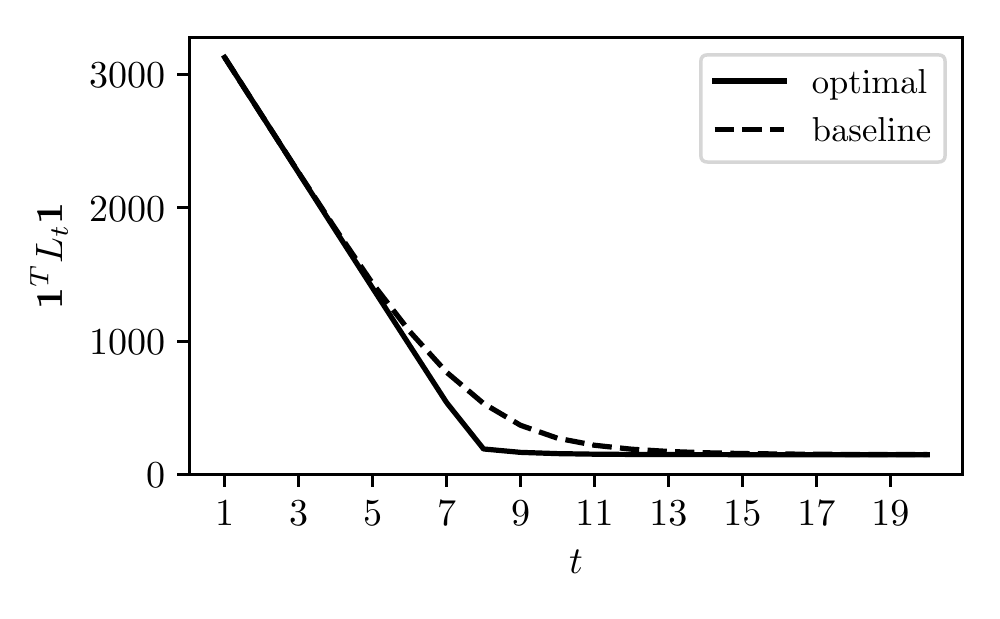}
  \caption{Total gross liability.}
\end{subfigure} %
\begin{subfigure}{0.45\textwidth}
\centering
  \includegraphics[width=\linewidth]{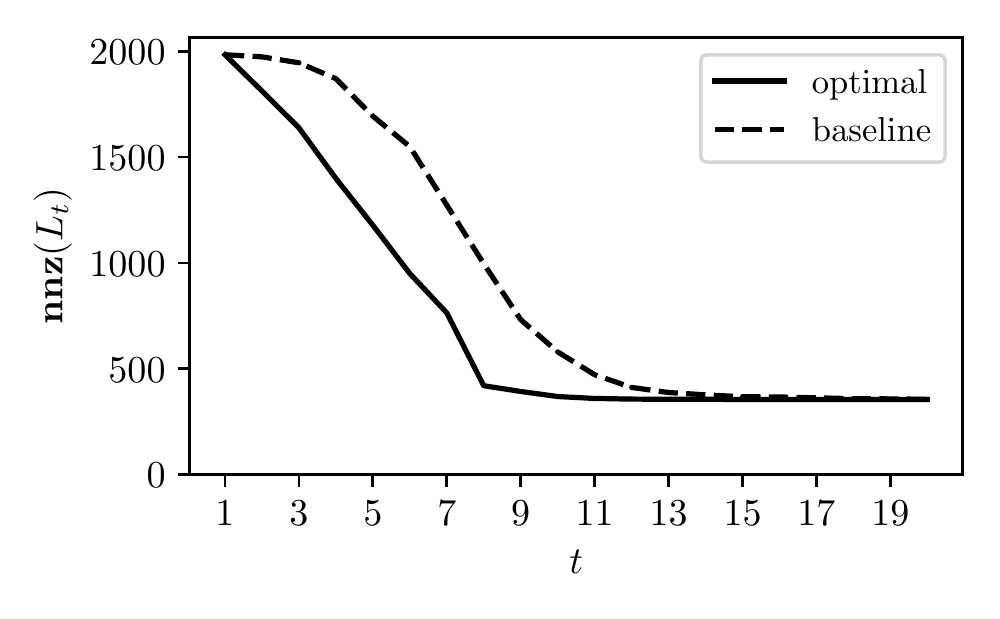}
  \caption{Number of non-cleared liabilities.}
\end{subfigure} %
\caption{Liability reduction example.}
\label{fig:4}
\end{figure}

Suppose that some entities have negative initial net worth.
This means that we will not be able to clear all
of the liabilities; our goal is then to reduce the liabilities as
much as possible, subject to the pro-rata constraint
$\diag(L^\mathrm{init}\ones) L_T = \diag(L_T\ones) L^\mathrm{init}$.
We consider the same liability matrix as \S\ref{sec:liability_clearing_example},
but change the initial cash to
\BEQ
c_1 = \max(L^\mathrm{init}\ones - (L^\mathrm{init})^T \ones + z, 0), \quad z_i \sim
\mathcal U(-5, 5), \quad i=1,\ldots,n,
\label{eq:not_enough_cash}
\EEQ
where $\mathcal U(-5,5)$ is the uniform distribution
on $[-5,5]$,
which in our case leads to 49 entities with negative net worth.
We consider the stage costs
\[
g_t(c_t,L_t,P_t)
=
\begin{cases}
\ones^TL_t\ones & P_t\ones \leq c_t/2, \\
+\infty & \text{otherwise},
\end{cases}
\quad g_T(c_T,L_T) = \ones^TL_T\ones.
\]
The stage cost is the total gross liability, plus
the indicator function of the constraint that each entity
pays out no more than half of its available cash in each time period.
We adjust the pro-rata baseline to
\[
P_t = \min(\diag(c_t/2)\Pi, L_t),
\]
so that each entity pays no more than half its available cash,
and increase the time horizon to $T=20$.
The results are displayed in figure \ref{fig:4}.
The optimal scheme is able to reduce the liabilities
faster than the baseline; both methods clear all
but around 350 of the original 2000 liabilities.
(In \S\ref{sec:num_entitites} we will see an extension that directly
includes the number of non-cleared liabilities in the stage cost.)

\subsection{Exogenous unknown inputs}
\begin{figure}
\centering
\includegraphics[width=.7\textwidth]{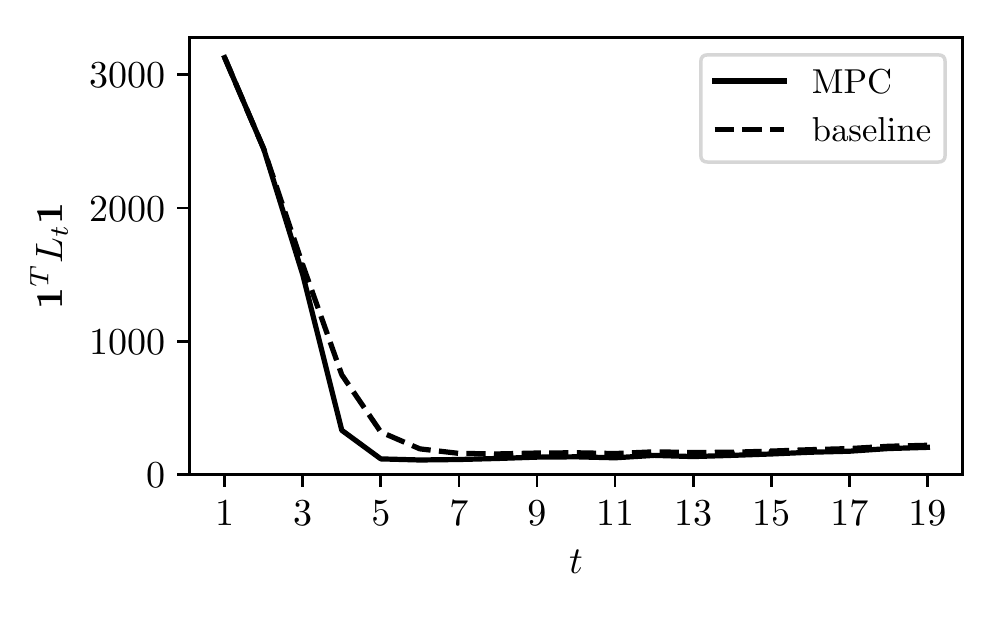}
\caption{Liability control with exogenous inputs.}
\label{fig:mpc}
\end{figure}

Next we consider the case where there are
exogenous unknown inputs to the dynamics.
The cash flows and change in liabilities are sampled according to
\[
w_t = z_1, \quad (W_t)_{ij} = \begin{cases}
z_2 / 10 & (L^\mathrm{init})_{ij}>0,\\
0 & \text{otherwise},
\end{cases} \quad t=1,\ldots,T-1,
\]
where $\log(z_1) \sim \mathcal N(0,I)$ and $\log (z_2)\sim\mathcal N(0,1)$.
At each time step $t$, we use the mean of the future inputs
as the forecast, or
\[
\hat w_{\tau \mid t} = e^{1/2}\ones, \quad
\hat W_{\tau\mid t} = \begin{cases}
e^{1/2}/10 & (L^\mathrm{init})_{ij}>0,\\
0 & \text{otherwise},
\end{cases} \quad \tau=t+1,\ldots,T-1.
\]

We sample the initial cash vector according to
\[
c_1 = \max(L^\mathrm{init}\ones - (L^\mathrm{init})^T \ones + z, 0), \quad z_i \sim
\mathcal U(-5, 0), \quad i=1,\ldots,n.
\]
We use the stage cost function $g_t(c_t,L_t,P_t) = \ones^TL_t\ones$
and the MPC policy described in \S\ref{sec:mpc}.
We compared the shrinking horizon MPC policy with the (modified)
pro-rata baseline policy described in \S\ref{sec:mpc}.
The results are displayed in figure \ref{fig:mpc};
note that the total gross liability appears to reach
a statistical steady state and the liabilities can never be fully cleared.
The MPC policy appears to be better than the baseline at reducing liabilities.

\section{Extensions and variations}
\label{s-extensions}

In this section we mention some extensions and variations on the 
formulations described above.

\subsection{Bailouts}
\label{s-bailouts}
We can add an additional term to the dynamics that injects 
cash into the entities at various times, with presumably very high cost 
in the objective.
With a linear objective term with sufficiently high weight, the bailout
cash injections are zero, if it is possible to clear the liabilities
without cash injection.
We note that bailouts have been considered in \cite[\S2.2]{capponi2015systemic}.

\subsection{Minimum time to clear liabilities}
Instead of the time-separable cost function given in \eqref{eq:liability_control},
we take as the objective the number of steps needed to clear all
liabilities.
That is, our objective is $T^\mathrm{clr}$, defined as the minimum value of $t$
for which $L_t=0$ is feasible.
It is easily shown that $T^\mathrm{clr}$ is a quasi-convex function of
the liability sequence $L_1, \ldots, L_T$ \cite[\S4.2.5]{boyd2004convex},
so this problem is readily solved using bisection, solving
no more than $\log_2 T$ convex problems.
If the liabilities cannot be cleared in up to $T$ steps
then we can find a $T$ such that they can be cleared
using the techniques described in \cite[\S3]{agrawal2020disciplined}.

\subsection{Non-time-separable cost}

The cost function in our basic formulation \eqref{eq:liability_control} is 
separable, \ie, a sum of terms for each $t=1, \ldots, T$.
This can be extended to include non-separable cost functions.
We describe a few of these below.  They are convex, but non-convex versions 
of the same objectives can also be employed, at the cost of computational
efficiency to solve the problem globally.

\paragraph{Smooth payments.}
Adding the term $\sum_{t=2}^{T-1}\|P_t - P_{t-1}\|_F^2$
to the cost,
where $\|A\|_F$ is the Frobenius norm, \ie, the square root
of the sum of the squared entries of $A$,
causes smooth transitions in the payment matrix.
This cost is sometimes called the Dirichlet energy \cite[\S7.3]{boyd2018introduction}
or a Laplacian regularization term \cite{biemond1990iterative}.

\paragraph{Piecewise constant payments.}
Adding the term $\sum_{t=2}^{T-1}\|P_t - P_{t-1}\|_1$,
where $\|A\|_1$ is the sum of absolute values
of the entries of $A$,
to the cost
encourages the payment matrix to change
in as few entries as possible between time steps.
This cost is sometimes called the total variation penalty \cite{rudin1992nonlinear}.

\paragraph{Global payment restructuring.}
Adding the term $\sum_{t=2}^{T-1}\|P_t - P_{t-1}\|_F$
to the cost encourages the entire payment matrix to change
at as few time steps as possible \cite{danaher2014joint}.

\paragraph{Per-entity payment restructuring.}
Adding the term $\sum_{t=2}^{T-1}\sum_{i=1}^n \|P_t^Te_i - P_{t-1}^Te_i\|_2$,
where $e_i$ is the $i$th unit vector,
to the cost
encourages the rows of the payment matrix,
\ie, the payments made by each entity,
to change at as few time steps as possible.
This penalty is sometimes called a group lasso penalty \cite{yuan2006model}.

\subsection{Infinite time liability control}
In \S\ref{sec:mpc} we described what is often called shrinking horizon control,
because at time period $t$, we solve for a sequence of 
payments $P_t, \ldots, P_{T-1}$
over the remaining horizon; the number of payments we optimize over (\ie,
$T-t$) shrinks as $t$ increases.
This formulation assumes there is a fixed horizon $T$.

It is also possible to consider a formulation with no fixed horizon $T$;
the liability clearing is done over periods $t=1,2, \ldots$ without end.
The exogenous inputs $w_t$ and $W_t$ also continue without end.
Since we have exogenous inputs, we will generally not be able to clear
the liabilities; our goal is only to
keep the liabilities small, while making if possible small payments.  
In this case we have a traditional infinite horizon control or regulator problem.
In economics terms, this is an equilibrium payment scheme.

The MPC formulation is readily extended to this case, and is sometimes
called receding horizon control (RHC), since we are always planning out $T$ steps
from the current time $t$.
It is common to add
a clearing constraint at the horizon in infinite time MPC or 
RHC formulations \cite[\S2.2]{rawlings2009model}.

\subsection{Minimizing the number of non-cleared liabilities}
\label{sec:num_entitites}
Another reasonable objective to consider is
the number of non-cleared (\ie, remaining) liabilities.
In this case, the only cost is the number of nonzero entries
in $L_T$.
This problem is non-convex, but it can be readily formulated as a mixed-integer convex
program (MICP), and solved, albeit slowly, using standard MICP
techniques such as branch-and-bound \cite{10.2307/1910129}.
It can also be approximately solved much quicker using heuristics,
such as iterative weighted $\ell_1$-minimization \cite{candes2008enhancing}.

As a numerical example, we consider a smaller version of the
initial liability matrix used in \S\ref{s-examples},
with $n=40$ and 400 nonzero initial liabilities.
We sample the initial cash according to \eqref{eq:not_enough_cash},
so that the liabilities cannot be fully cleared,
and use a time horizon $T=10$.
Minimizing the sum of total gross liabilities
takes 0.05 seconds, resulting in 46 non-cleared liabilities
and a final total gross liability of 22.52.
By contrast, minimizing the number of non-cleared liabilities
takes 22.93 seconds,
resulting in only 10 non-cleared liabilities
and a final total gross liability of 29.78.
(The increase in computation time of a mixed-integer
convex optimal control problem, compared to a convex optimal control 
problem of the same size, increases rapidly with problem size.)

As an extension of minimizing the number of 
non-cleared liabilities, we can consider minimizing the 
number of non-cleared entities.  If the $i$th row of $L_t$ is zero,
it means that entity $i$ does not owe anything to the others, and we 
say this entity is cleared.  We can easily add the number of non-cleared
entities to our stage cost, using a mixed-integer convex formulation.

\subsection{Distributed algorithm}
As stated, the liability control problem \eqref{eq:liability_control}
requires global coordination, \ie, full knowledge of the
cash held and the liabilities between the entities
throughout the optimization procedure.
In many settings where cash, liabilities, or payments
cannot be publicly disclosed, this is not possible.

It is possible to solve the liability control problem
in a distributed manner where each
entity only knows its cash
and the payments and liabilities it is involved in
during the optimization procedure.
That is, entity $i$ only needs to know $(c_t)_i$,
the $i$th row and column of $L_t$,
and the $i$th row and column of $P_t$.

We can do this by adding a variable
$\tilde P_t\in\reals_{+}^{n \times n}$,
the constraint $\tilde P_t = P_t^T$,
$t=1,\ldots,T$,
and replacing the cash dynamics \eqref{e-cupdate}
with
\[
c_{t+1} = c_t - P_t \ones + \tilde P_t\ones, \quad t=1,\ldots,T-1.
\]
Then, by applying the alternating direction method
of multipliers (ADMM) to the splitting $(c_t,L_t,P_t)$ and $\tilde P_t$,
we arrive at a distributed algorithm for the problem \cite{boyd2011distributed}.
Each iteration of the algorithm involves three steps;
1) each entity solves a separate control problem
to compute their cash, outbound liabilities, and outbound payments;
2) each entity solves a separate least squares problem that depends
on their inbound payments; and 3) each entity performs a separate
dual variable update. When the stage cost is convex,
this algorithm is guaranteed to converge to a (global)
solution \cite[Appendix A]{boyd2011distributed}.
Each step of the algorithm only requires coordination
between entities connected in the liability graph,
and hence preserves some level of privacy.
Similar ideas have been used to develop
distributed privacy-preserving implementations of
predictive patient models across hospitals \cite{jochems2016distributed}
and energy management across microgrid systems \cite{liu2017distributed}.

\section*{Acknowledgements}
The authors would like to thank Zachary Feinstein
for his helpful discussion and comments, in particular for
providing many useful references and
coming up with the idea of a distributed implementation.
The authors would also like to thank Daniel Saedi
for general discussions about banking.
Shane Barratt is supported by the National Science Foundation Graduate Research Fellowship
under Grant No. DGE-1656518.

\clearpage

\bibliography{refs}

\clearpage

\appendix

\section{Sparsity preserving formulation}
\label{s-form2}
In this section we describe a sparsity-preserving formulation
of problem \eqref{eq:liability_control}.
We make use of the fact that
$L_t$ and $P_t$ are at least as sparse as $L^\mathrm{init}$
(see \S\ref{s-setup}).

First, let $m=\mathbf{nnz}(L^\mathrm{init})$
and $I_k \in \{1,\ldots,n\} \times \{1,\ldots,n\}$, $k=1,\ldots,m$,
be the sparsity pattern of $L^\mathrm{init}$, meaning
$(L^\mathrm{init})_{ij}=0$ for all $(i,j)\not\in I_k$, $k=1,\ldots,m$.
Instead of working with the matrix variables $L_t$ and $P_t$,
we work with the vector variables $l_t\in\reals_+^m$
and $p_t\in\reals_+^m$,
which represent the nonzero entries of
$L_t$ and $P_t$ (in the same order).
That is,
\[
(l_t)_k = (L_t)_{ij}, \quad (p_t)_k = (P_t)_{ij}, \quad (i,j) = I_k, \quad k=1,\ldots,m.
\] 
The initial liability is given by $l^\mathrm{init}\in\reals_+^m$,
which contains the nonzero entries of $L^\mathrm{init}$.
The sparsity preserving formulation
of the optimal control problem \eqref{eq:liability_control} has the form
\BEQ
\begin{array}{ll}
\mbox{minimize} & \sum_{t=1}^{T-1} g_t(c_t,l_t,p_t)  + g_T(c_T,l_T)\\
\mbox{subject to} & l_{t+1} = l_t - p_t, \quad t=1,\ldots,T-1,\\
& c_{t+1} = c_t - S^\mathrm{row}p_t + S^\mathrm{col} p_t, \quad t=1,\ldots,T-1,\\
& S^\mathrm{row} p_t \leq c_t, \quad t=1,\ldots,T-1,\\
& p_t \geq 0, \quad t=1,\ldots,T-1,\\
& l_t \geq 0, \quad t=1,\ldots,T,\\
& c_1 = c^\mathrm{init}, \quad l_1 = l^\mathrm{init}, \quad c_T \geq 0,
\end{array}
\label{eq:liability_control_sparse}
\EEQ
where $S^\mathrm{row}\in\reals^{n \times m}$
sums the rows of $P_t$, \ie,
$S^\mathrm{row}p_t=P_t\ones$,
and $S^\mathrm{col}\in\reals^{n \times m}$
sums the columns of $P_t$, \ie,
$S^\mathrm{col}p_t=P_t^T\ones$.
The cost functions are applied only to the nonzero entries
of $L_t$ and $P_t$, so they
take the form $g_t:\reals_+^n \times \reals_+^m \times \reals_+^m\to\reals\cup\{+\infty\}$
and $g_T:\reals_+^n \times \reals_+^m\to\reals\cup\{+\infty\}$.
Problem \eqref{eq:liability_control_sparse}
has just $2T(n+m)$ variables, which can be much fewer
than the original $2T(n+n^2)$ variables when $m \ll n^2$.

\end{document}